\documentclass[prl,onecolumn,showpacs,preprintnumbers,amsmath,amssymb]{revtex4}
\usepackage{amsfonts,amsmath}
\usepackage{graphicx}
\usepackage{dcolumn}

\newcommand {\pd}{\partial}
\newcommand {\beq}{\begin{equation}}
\newcommand{\eeq}{\end{equation}}
\newcommand {\bseq}{\begin{subequations}}
\newcommand{\eseq}{\end{subequations}}
\newcommand{\bal}{\begin{align}}
\newcommand{\eal}{\end{align}}

\newcommand{\lm}{\lambda}
\newcommand{\al}{\alpha}

\newcommand{\gm}{\gamma}

\newcommand{\s}{\sigma}

\newcommand{\ep}{\varepsilon}

\newcommand{\dphi}{\phi^{\dagger}}

\begin{document}
\setlength{\unitlength}{1mm}

   \title{Vicious L\'evy flights }
  \author {Igor Goncharenko and Ajay Gopinathan }
  \affiliation{School of Natural Sciences, University of California, Merced, California, 95343, USA}
    \begin{abstract} 
    We study the statistics of encounters of L\'evy flights by introducing the concept of vicious L\'evy flights - distinct groups of walkers performing independent L\'evy flights with the process terminating upon the first encounter between walkers of different groups. We show that the probability that the process survives up to time $t$ decays as $t^{-\al}$ at late times. We compute $\al$ up to the second order in $\ep$-expansion, where $\ep=\s-d$, $\s$ is the L\'evy exponent and $d$ is the spatial dimension. For $d=\s$, we find the exponent of the logarithmic decay exactly. Theoretical values of the exponents are confirmed by numerical simulations.      
   \end{abstract}
  \maketitle
  
  Diffusive processes with long range jumps play an important role in many physical, chemical and biological phenomena. A L\'evy flight is an example of such a process where the probability distribution of the length of an individual step, $r$, is governed by the power-law $r^{-d-\s}$, where $d$ is the dimension of the space and $\s$ is the L\'evy exponent. Smaller values of $\s$ therefore produce longer range jumps while for $\s\ge2$, the mean jump length is finite and simple diffusive behavior is recovered. L\'evy flights have been used to describe a wide range of processes including epidemic spreading, transcription factor proteins binding to  DNA, kinetic Ising models with long-range interactions, foraging animals  and light propagation in disordered optical materials \cite{Klaft2,DNA,MenyOdor, Berg, VishNat,Barth}. While individual L\'evy flights have been studied in great detail, the same is not true if we consider several distinct groups of L\'evy flights. One could, for example, be interested in the statistics of encounters between members of different groups. This question is relevant for processes where the outcome depends on the occurrence of such encounters. Examples include sharks and other marine animals searching for prey \cite{Sims}, chemical reactions in turbulent environments \cite{Klaft1}, electron-hole recombination in disordered media \cite{Shles} and even male spider-monkeys encountering their mates or other aggressive males in the forest \cite{Ramos}. \par
  In this Letter we compute the survival probability, i.e. the probability that no two members of different groups of L\'evy flights have met up to time $t$. For the case of simple diffusion with exactly one particle in each group, this corresponds to the classic problem of Gaussian vicious walks \cite{Fisher}, i.e. walks that are prohibited from being on the same site at the same time, but remain independent otherwise. Here we generalize this concept to groups of L\'evy flights under the same constraints. We term them vicious L\'evy flights (VLF). We consider $p$ sets of particles with $n_i$ particles in each set, $i=1\dots p$,  that are driven by L\'evy noise on the $d$ dimensional regular lattice. A pairwise interset  short-range (delta-function) interaction is introduced to guarantee that trajectories which continue beyond an intersection are discarded, i.e. have zero statistical weight. This terminates the process at the first encounter between members of different groups. Particles belonging to the same set do not interact.  We note that L\'evy  flights are allowed to jump over each other, unlike ordinary random walks which can only jump to neighboring sites and can not intersect with the vicious constraint. In $d=1$ this means that the ordering is preserved for vicious walks but not for VLF. For simplicity we assume that  L\'evy exponents for all flights are the same. Generalization to the case of different L\'evy exponents will be done elsewhere. At time $t=0$  all particles start in the vicinity of the origin. We are interested in the survival probability of this system at late times.

We start with a field theoretic formulation of the problem. Methods to formulate field theories for such stochastic systems are well established \cite{Doi,Peliti}. Specifically for Gaussian vicious walks, such a formulation exists \cite{Cardy} and the form of the action is known. We can adapt the action to our case by replacing the Laplacian $\nabla^2$ with the operator $\nabla^{\s}$ that generates long-range jumps. This gives
\beq\mathcal{S} (\phi_i,\dphi_i) = \int dtd^dx \sum_{i=1}^p [ \dphi_i\pd_{t} \phi_i + \dphi_i\nabla^{\s} \phi_i]    + \sum\limits_{1\le i<j\le p} \lm_{ij} \dphi_i(t,x)\phi_i(t,x)\dphi_j(t,x)\phi_j(t,x). \label{hamil}\eeq
where $\phi_i(x,t)$ are $p$ complex order parameters corresponding to $p$ different sets of equivalent L\'evy flights and  $\lm_{ij}$ are coupling constants corresponding to interset interactions. The non-intersection property of VLF arises from the choice $\lm_{ij}\to\infty$ but we will show that to leading order the survival probability does not depend on the particular value of these coupling constants. This action is also similar to the action for the reaction-diffusion problem with long-range interactions \cite{Jans, Hinr}. 
Power counting shows that the upper critical dimension for the above field theory is $d_c = \s$ for $\s<2$ and $d_c=2$ for $\s\ge2$. VLF exhibit different phases (see inset Figure 1) depending on the values of $d$ and $\s$. In the mean field phase for $d>d_c$ (Region I), the survival probability of VLF is non-zero at infinite time because the walks become non-recurrent and particles can avoid each other for all time. For $\s\ge 2$ (Region II) VLF reproduce Gaussian vicious walks. For $d<\s<2$ (Region III) we expect fluctuations to play an important role. In this phase, we will obtain the critical behavior of the survival probability using $\ep$-expansion  ($\ep=\s-d$) around mean field theory in two-loop approximation. \par
    \begin{figure}
\includegraphics[scale=0.40]{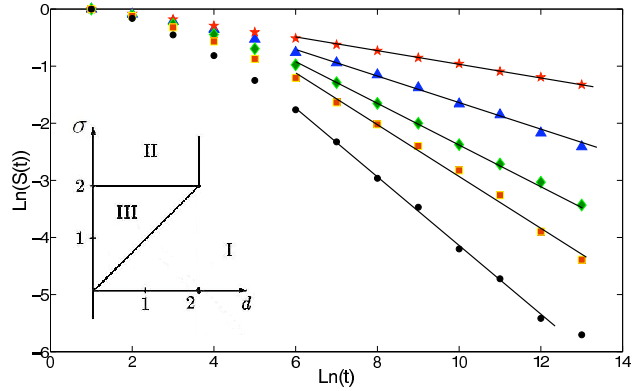}
\vspace{-0.1in}
\caption{\label{fig:1014} a.  $\ln(S)$ vs $\ln(t)$ for two identical VLF in $d=1$. $\s$ values from top tp bottom are $1.1, 1.3, 1.5, 1.7, 2.5$ respectively. Symbols represent simulation data and solid lines correspond to best fit lines to the late time data. Inset: Domains of VLF exponents in the $\s-d$ plane.}
\vspace{-0.15in}
\end{figure}

We now turn to the renormalization group analysis. The propagator given by (\ref{hamil}) is $\Gamma^{(1,1)}_{j}(s,k) = (s+k^{\s})^{-1}$. The particular form of the  vertex in (\ref{hamil}) leads to the fact that there are no diagrams which dress the propagator. This implies that the bare propagator  is the exact propagator for the theory. The proper vertex is defined by factoring out external legs  from the  ordinary 4-point Green's function of (\ref{hamil}) $G^{(2,2)}_{ij}(s_l,k_l;\lm)$. Here $\lm= \{\lm_{ij}\}$ is the collection of coupling constants and $(s_l, k_l)$ for $l=1\dots 4$ are the energy (Laplace image of time) and momenta respectively \cite{Bronzan}. This yields 
\beq \Gamma^{(2,2)}_{ij}(s_l, k_l;\lm) = \frac{G^{(2,2)}_{ij}(s_l,k_l;\lm)}{\prod_{m=1}^4 \Gamma^{(1,1)}(s_m,k_m)}.\eeq  
The renormalized coupling, $\lm_{Rij}$, is the value of the proper vertex at $(s_l=\mu, k_l=0),$ for all $l$, where $\mu$ is a renormalization group flow scaling parameter. It is possible to sum all diagrams in the series  with the result \beq\label{lmR}\lm_{Rij} = \lm_{ij}(1+\lm_{ij} I_1)^{-1},\eeq where $I_1 = (2\pi)^{-d}\int d^dq(2\mu +2q^{\s})^{-1}$. The value of the integral $I_1$  is given by $I_1 = A \mu^{-\ep/\s}\ep^{-1}$, where 
  \beq\label{A} A=\frac{\ep\Gamma(d/\s)\Gamma(\ep/\s)}{2^{d} \pi^{d/2} \s \Gamma(d/2)} = \frac{2^{-\s}\pi^{-\s/2}}{\Gamma(\s/2)} + O(\ep),\eeq is the geometric factor at the leading order in $\ep=\s-d$. By introducing the redefined  coupling constant $g_{Rij} =\lm_{Rij} \mu^{-\ep/\s}$, we obtain the following renormalization group flow equations (see Appendix A for details): \beq\label{beta}\mu \frac{\pd g_{Rij}}{ \pd \mu} = (-\ep + A g_{Rij})g_{Rij}/\s.\eeq The fixed point of this flow is  $g_* = \ep A^{-1}$. We note that this value of the fixed point is exact to all orders since all diagrams were taking into account in (\ref{lmR}). The stability of the fixed point follows from the fact that $-\pd\beta_{ij}/\pd g_{Rij}|_{g_{Rij} =\ep /A } = -\ep/\s < 0$ in the VLF region, where $\beta_{ij}=\mu \pd g_{Rij} /\pd \mu$ is renormalization group beta function. 
 
 \begin{figure}
\includegraphics[scale=0.40]{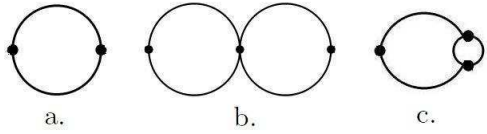}
\vspace{-0.15in}
\caption{\label{fig:fdVLF} a.  1-loop diagram corresponding to the integral $I_1$ b.,  c.  2-loop intgrals corresponding to $I_1^2$ and $I_2$ respectively.}
\vspace{-0.15in}
\end{figure}

 We  now consider the survival probability  which is defined as the correlation function \cite{Cardy} 
 \begin{equation}
\label{sp-G}
 S(t;\lm)
=\int\prod_{i=1}^p\prod_{\alpha_i=1}^{n_i}d^dx_{i,\alpha_i}
\langle\phi_i(t,x_{i,\alpha_i})(\dphi_i(0,0))^{n_i}\rangle,
\end{equation}
 with the measure $\int{\cal D}\dphi{\cal D}\phi\exp[-\mathcal{S}]$. The Feynman diagram of (\ref{sp-G}) at zero order is  a vertex with $2N$ external legs. Similar to the case of the $(2,2)$-vertex it is convenient to work with the truncated correlation function $ \Gamma(s_l, k_l;\lm) = S(s_l,k_l;\lm)/\prod^{2N}_{m=1}\Gamma^{(1,1)}(s_m,k_m)$. The finite renormalized truncated correlation function $\Gamma_{R}(s_l,k_l;\lm_R,\mu)$, where $\lm_R=\{\lm_{Rij}\}$ is a collection of renormalized coupling constants,  is related to the  bare truncated correlation function by $\Gamma_{R}(s_l,k_l;\lm_R,\mu)=Z(\lm,\mu)\Gamma(s_l,k_l;\lm)$, where $Z(\lm,\mu)$ is the scaling function. From this one obtains the renormalization group equation for  $\Gamma_{R}(s_l,k_l;\lm_R,\mu)$ using the chain rule
 \beq\label{CS} (\mu\frac{\pd}{\pd\mu} + \beta_{ij}\frac{\pd}{\pd g_{Rij}} + \gm)\Gamma_{R}(s_l,k_l;\lm_R,\mu)=0, \eeq 
 where $\gm = \mu\pd\ln Z/\pd\mu.$ At the fixed point (\ref{CS}) reduces to $(\pd/\pd\ln(\mu) +  \gm_*)\Gamma_{R}(s,\mu)=0$
 whose solution is \beq\label{solCS} \Gamma_R \sim \exp(\int_0^\mu \gm_* d(\ln(\mu'))),\eeq where $\gm_*=\mu\pd\ln Z/\pd\mu|_{g_{Rij}=g_*}$. Since $\gm_*$ is constant at the fixed point we have $\Gamma_R \sim \mu^{-\gm_*} $. The fact that the dimensions of field and the action are $[\dphi]=[\phi]=k^{d/2}$ and $[\mathcal{S}]=1$ implies that $[S]=1$. Thus it follows that the  survival probability can only be a function of the dimensionless product $\mu t$.
 From this one  infers that the asymptotic behavior of the survival probability  is $S(t)\sim t^{-\gm_*}$ which gives $\al=\gm_*$. In order to find $Z$ one uses a normalization condition on $\Gamma_{R}$ that fixes the value of $Z$. This can be chosen as  $\Gamma_{R}(s_l,k_l;\lm_R,\mu)=1$ when $s_l=\mu$ and $k_l =0$ for all $l$. This implies that $Z=\Gamma(\mu,0;\lm)^{-1}$. $\Gamma(\mu,0;\lm)$ can be expressed as a series, up to two-loop order, of the integrals corresponding to the diagrams shown in Fig.2 with appropriate combinatorial factors originating from the number of distinct ways in which propagators can be assigned to the same diagram.  $I_1$ was evaluated before. The integral corresponding to the diagram 2c is
\beq\label{I2} I_2 = \int \frac{d^dkd^dq}{(2\mu + 2k^{\s})(3\mu + q^{\s}+k^{\s}+|k+q|^{\s})}\eeq
 This can be evaluated using Mellin-Barnes representation \cite{Smir}  which replaces the sum in the denominator of $|k+q|^{\s}$ and $q^{\s}$ by the product of  these terms raised to some power. The result for $I_2$ up to the leading order reads 
 \beq\label{I2ep} I_2 = \frac{2^{-2\s}\pi^{-\s}}{\Gamma(\s/2)^2} \mu^{-2\ep/\s} \left( \frac{1}{2\ep^2} + 
 \frac{2(-\log(3/4)/4 + C)}{\s\ep} \right),\eeq
where $C= [\psi^{(0)}(\s/2) + \log(4\pi)]/2$ and $\psi^{(0)}(x)$ is standard digamma function. The details of the calculations are summarized in  Appendix B. Knowing $I_1,I_2$ and the appropriate combinatorial factors allows us to evaluate $\Gamma(\mu,0;\lm)$ and therefore $Z(\mu,\lm)$ as a series. Differentiating $\ln Z$ with respect to $\mu$ and substituting $\lm$ with $\lm_R$ (inverting (\ref{lmR})) and then taking the value at the fixed point gives us the survival probability exponent $\al=\gm_*$, with the final result (see Appendix C):
 \beq\label{SPexp} \al = \sum_{1\le i<j\le p}n_in_j \ep/\s + \ln(3/4)Q\ep^2/\s^2 , \eeq
where $Q =  6\sum_{1\le i<j<k\le p} n_in_jn_k +  \sum_{1\le i<j\le p}n_in_j(n_i+n_j-2)$.
At the critical dimension $d=d_c=\s$ we see that the fixed point coincides with the Gaussian point. 
The interaction becomes marginal in the renormalization group sense. Equation (\ref{beta}) then yields the flow equations for the running coupling constant \beq x d\bar{g}_{ij}(x)/dx =  A \bar{g}^2_{ij}(x)/\s,\eeq with initial condition $\bar{g}_{ij}(1)=g_{Rij}.$ Solving this and substituting the result into  (\ref{solCS}) we get  $ S(t) = (\ln t)^{-\al_l}, $ where \beq \al_l = \sum_{1\le i<j\le p} n_i n_j . \eeq

\begin{figure}
\includegraphics[scale=0.60]{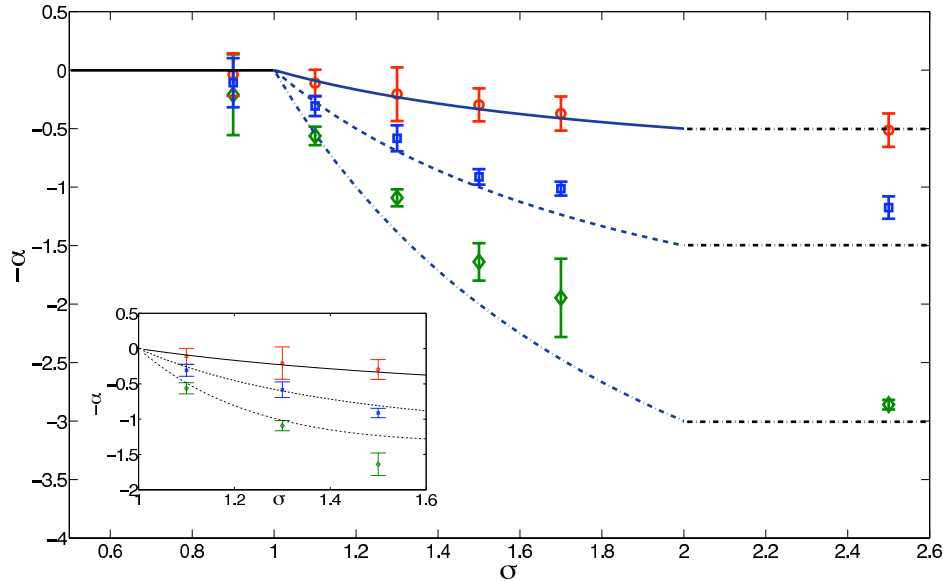}
\vspace{-0.15in}
\caption{\label{fig:exp235}  $\al$ as a function of $\s$ for $d=1$. Symbols with error bars represent simulation data corresponding from top to bottom to $N=2,3,4$ VLF respectively. Lines correspond to 1-loop approximation from formula (\ref{SPexp}) for $1<\s <2$. For $\s<1$, $\s\ge2$ lines represent the mean field and Gaussian exponents respectively. Inset: Same simulation data compared to 2-loop approximation from (\ref{SPexp})}
\vspace{-0.15in}
\end{figure}

\begin{figure}
\includegraphics[scale=0.40]{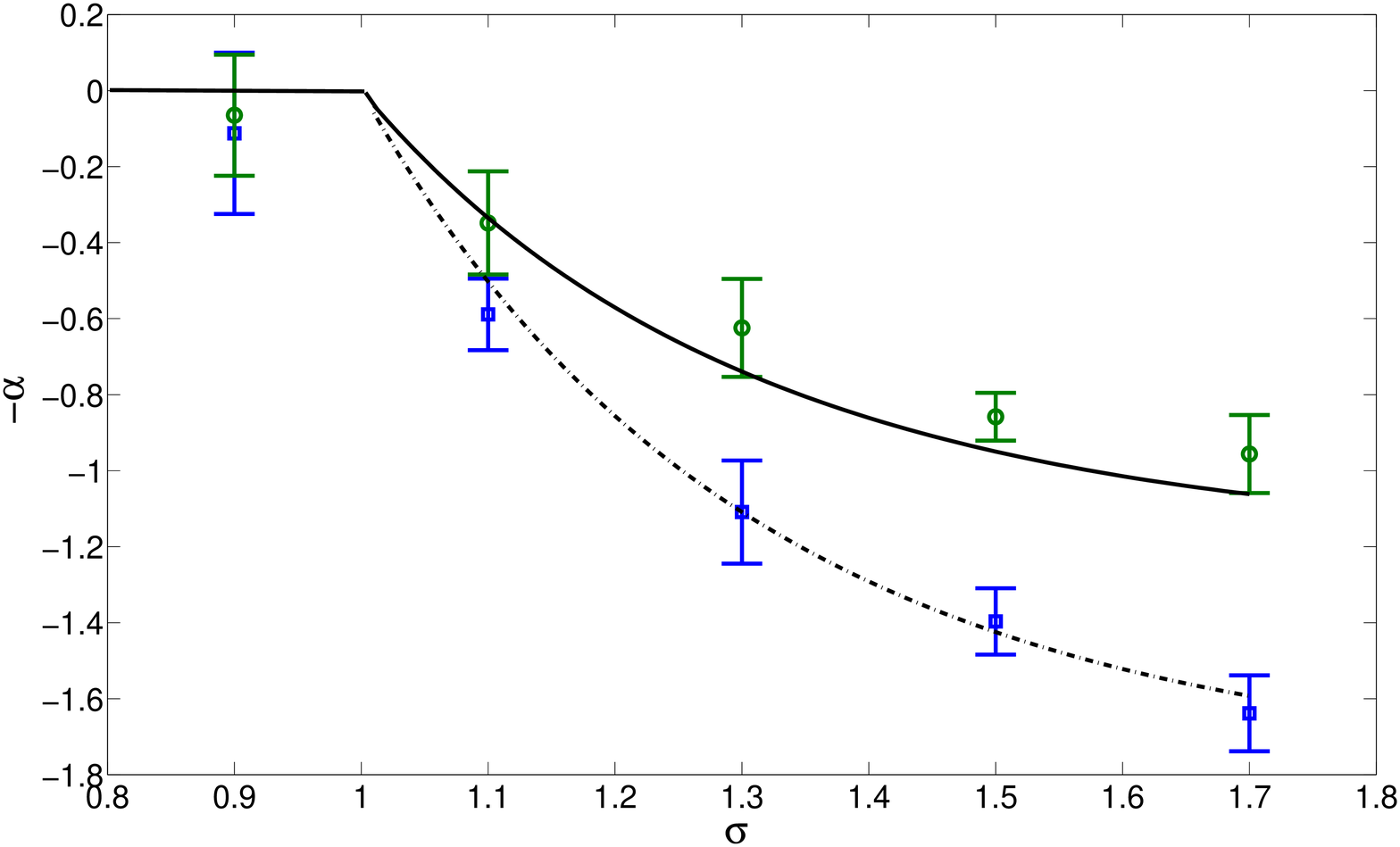}
\vspace{-0.35in}
\caption{\label{fig:LL} $\al$ vs $\s$ for predator and prey problem in $d=1$. Symbols represent simulation data for 4 predators and 1 prey (circles) and 3 predators and 2 prey (squares). Lines are 2-loop approximation from formula (\ref{SPexp}).}
\vspace{-0.2in}
\end{figure}

Now we describe the details of the numerical simulation that we used to confirm our results in $d=1$. At $t=0$ we start with $N=\sum_{i=1}^p n_i$ particles belonging to $p$ distinct sets placed equidistantly on the lattice. At each time step we generate  $N$ 
random variables, $x_j$, drawn from the uniform distribution on the interval $(0,1)$. Each particle jumps a distance $l_j = x_j^{-1/\s}$ with equal probability to the left or to the right. This procedure generates an independent L\'evy flight trajectory for each particle. The process stops whenever particles from different sets land on the same site. We perform $\sim 10^5$ iterations for each set of parameter values. The survival probability $S(t)$ is defined by the number of processes that survived beyond time $t$ divided by the total number of iterations. Figure 1 shows the plot of the survival probability as a function of time for $N=2$ for different values of $\s$. It is clear that at late times $\ln S(t)$ is linear in $\ln t$ verifying our predicted power-law decay $S(t)\sim t^{-\al}$. The critical exponent $\al$ is evaluated from the slope of the best fit line of the late time data. 

We first consider systems with exactly one particle in each set. Figure 3 shows the value of exponent $\al$ for various values of $\s$ with the total number of particles $N=2,3$ and $4$ in $d=1$. Values of $\s\ge 2$ will reproduce Gaussian vicious walks and we therefore expect our exponents to approach the exact Fisher exponents \cite{Fisher} as seen in  Figure 3. For two VLF higher loop corrections are absent (see (\ref{SPexp})) and the one-loop result is an exact result in agreement with the simulation. It is interesting to note that the survival probability for the $N=2$ case is equivalent to the first return probability of a single L\'evy flight to the origin after time $t$ which scales as $t^{-1+d/\s}$ \cite{Klaft3} and matches our results. For $\s<1$, or $d>d_c$,  we expect mean field behavior where survival probability at late times approaches a non-zero value implying that there is a finite probability that L\'evy flights with $\s<1$ will never find each other. The fact that the survival probability tends to a constant at late times is reflected in the small values of $\al$ for $\s=0.9$. For  $1<\s<2$, the mean field behavior is incorrect and we expect the fluctuations to shift the decay exponent to some non-trivial value. For $\s$  close to one, the 1-loop result is in good agreement with the simulation. For larger values of $\s$, the 2-loop corrections perform better (see Fig.3 inset). It is to be noted that the discrepancy between theory and simulation becomes large for higher values of $N$ because the combinatorial factors in (\ref{SPexp}) become large and we therefore need to keep higher order terms in $\ep$ for the same degree of accuracy. It is interesting that the 1-loop approximation works reasonably well over the entire range of $1< \s \le 2$ simply because the 1-loop term in (\ref{SPexp}) happens to give the exact Fisher result if we set $\s=2$. We notice that in all cases  the value of the survival probability exponent $\al$ increases with $\s$ starting from $\al\sim 0$ at $\s<1$ and rising to the value of the Fisher exponents for the equivalent Gaussian vicious walks. This is in contrast to diffusion-annihilation reactions with long-range jumps where the density of reactants decays faster for smaller values of $\s$ \cite{Hinr}. 

We now consider a system that consists of 2 sets of identical VLF with different numbers of independent particles in each set. We shall call one set predators and the other set prey. Figure 4 compares the values of the 2-loop exponents to the simulation results for various values of $\s$ for two different cases: 4 predators - 1 prey and 3 predators - 2 prey. Similar to the previous case we have mean field and Gaussian behaviors for $\s<1$  and $\s\ge 2$ respectively. The Gaussian case is also known as the lion-lamb problem and has been studied before \cite{Redner}. Unlike the lion-lamb problem, however, our results do not depend on the initial ordering of predators and prey because ordering is not preserved for VLF. For a given $\s$ and total number of predators plus prey, the number of potentially lethal encounters is maximized when the difference between the number of predators and prey is the smallest implying that the survival probability will decay faster as seen in Fig.4.

Our results suggest that it is interesting to solve the problem in the general case of particles with different diffusion constants and L\'evy exponents. The predictive power of the $\ep$-expansion for VLF, that we have demonstrated, should be useful in many applications  of practical importance.  Examples include the optimization of the predator-prey search \cite{Vish1} or trapping probabilities \cite{Hub}. Generalization to the case of intelligent predators, i.e. interacting with a prey by means of the long-range potential, may lead to different critical behavior \cite{Gonch,SatMaj1,SatMaj2}.  Simple diffusion processes in  power-law small world networks are effectively  L\'evy flights \cite{Kozma} with the exponent $\s$ controlling the distribution of long-range links. Our work could be used to understand what network structure, or what $\s$, would optimize the search and how much more efficient  several independent searchers will be.

  The authors would like to acknowledge UC Merced start-up funds and a James S. McDonnell Foundation Award for Studying Complex Systems.

\section{Appendix A}

Here we derive the 1-loop integral \beq I_1=I_1(\mu) = (2\pi)^{-d}\int d^dk (2\mu + 2k^{\s})^{-1}.\eeq We will use dimensional regularization. First we notice that there is no angle dependence under the integral thus one can integrate out $d-1$ angle variable and use alpha representation: 
\beq X^{-\lm} = \Gamma(\lm)^{-1}\int_0^{+\infty}d\alpha \alpha^{\lm-1} \exp(-\alpha X) \eeq
 to handle 1d momenta integral: \beq I_1 = \frac{S_d}{2}\int\limits_0^{+\infty} \frac{k^{d-1}dk}{\mu + k^{\s}}=\frac{S_d}{2}\int\limits_0^{+\infty}\int\limits_0^{+\infty}d\alpha dk k^{d-1}\exp(-\alpha\mu - \alpha k^{\s})= 
\frac{S_d}{2\s}\Gamma(d/\s)\int\limits_0^{+\infty}d\alpha \alpha^{-d/\s} \exp(-\alpha\mu), \eeq
 where $S_d = 2\pi^{d/2}/\Gamma(d/2)$ is the area of the d-dimensional unit sphere. After taking the integral over $\alpha$ one has \beq I_1 = \frac{S_d}{2\s}\Gamma(d/\s) \Gamma(\ep/\s) \mu^{-\ep/\s} = A \mu^{-\ep/\s} \ep^{-1} + O(\ep^0), \eeq
where $A$ has been defined by the formula (\ref{A}). 

Now we show details of deriving renormalization
group flow equations (\ref{beta}).  We start with equation (\ref{lmR}) and express $\lm_{ij}$ in terms of $\lm_{Rij}$. The result reads 
\beq \lm_{ij} = \frac{\lm_{Rij}}{1-\lm_{Rij}I_1}. \eeq
Multiplying left and right hand side of the last equation on $\mu^{-\ep/\s}$ and redefining the coupling constant $g_{Rij} =\mu^{-\ep/\s}\lm_{Rij}$ we infer that \beq\label{lm-gR} \mu^{-\ep/\s}\lm_{ij} = g_{Rij}/(1-g_{Rij}A\ep^{-1}).\eeq 
Differentiating left and right hand side of (\ref{lm-gR}) with $\mu\frac{\pd}{\pd\mu}$ we obtain
\beq\label{Dlm-gR}  (-\ep/\s) \mu^{-\ep/\s}\lm_{ij} = -\beta_{ij} g^{-2}_{Rij}/(g^{-1}_{Rij}-A\ep^{-1})^2 \eeq
Now we substitute (\ref{lm-gR}) into (\ref{Dlm-gR}) and find beta function up to second order in small $\ep$ and $g_{Rij}$ expansion: \beq \beta_{ij} = (-\ep g_{Rij} + Ag^2_{Rij})/\s + O(\ep g^2)\eeq

\section{Appendix B}

Here we derive the 2-loop integral \beq I_2=I_2(\mu)=(2\pi)^{-2d}\int d^dk d^dq [(2\mu + 2k^{\s})(3\mu +k^{\s}+q^{\s}+|k+q|^{\s})]^{-1}.\eeq 
The term $|k+q|^{\s}$ leads to the appearance of angle integration. Nevertheless it is possible to avoid angle integration. The key idea is to use Mellin-Barnes representation \cite{Smir}: \beq\frac{1}{(X+Y)^{\lm}} = \int_{-i\infty}^{+i\infty}\frac{dz}{2\pi i} \frac{Y^z}{X^{\lm+z}} \frac{\Gamma(\lm+z)\Gamma(-z)}{\Gamma(\lm)} \eeq
Applying MB formula twice we split the sum of two terms containing $q$ integration into the factor of these terms raised to some power: 
\begin{align} I_2 &=\int \frac{d^dk d^dq}{2(2\pi)^{2d}}\int\limits_{-i\infty}^{+i\infty}\frac{dz}{2\pi i}
\frac{\Gamma(1+z_1)\Gamma(-z_1)}{\mu + k^{\s}} \frac{(3\mu +k^{\s}+q^{\s})^{z_1}}{|k+q|^{\s (1+z_1)}} \nonumber\\
 &= 
\int \frac{d^dk d^dq}{2(2\pi)^{2d}}\int\limits_{-i\infty}^{+i\infty}\frac{dz_1dz_2}{(2\pi i)^2}
\frac{\Gamma(1+z_1)\Gamma(-z_1+z_2)\Gamma(-z_2)}{\mu + k^{\s}} \frac{(3\mu +k^{\s})^{z_2}}{|k+q|^{\s (1+z_1)}q^{\s (-z_1+z_2)}} 
\,, \label{I2zz}
\end{align}
Now integral over $q$ becomes standard: 
\beq\label{Iq} I_q= \int \frac{d^dq}{(q^2)^{a_1}((k+q)^2)^{a_2}} = \pi^{d/2} k^{d-2(a_1+a_2)}
 \frac{\Gamma(a_1+a_2 -d/2)\Gamma(d/2 - a_1)\Gamma(d/2 -a_2)}{\Gamma(a_1)\Gamma(a_2)\Gamma(d-a_1-a_2)},\eeq
where $a_1=\s(-z_1+z_2)/2$ and $a_2=\s(1+z_1)/2$. Thus we will be left with integral over $k$ of the form:
\beq I_k= \int \frac{d^dk k^{-\ep-\s z_2} (3\mu +k^{\s})}{2\mu+2k^{\s}} \eeq
The function under the integral does not depend on the angle and therefore $I_k$ can be cast into one dimensional integral over momenta:

\beq I_k = \frac{S_d}{2\s}\int\limits_0^{+\infty}dk k^{-2\ep/\s-z_2} \frac{(3\mu+k)^{z_2}}{\mu+k}\eeq
We will compute this integral using alpha representation.
\beq I_k  = \frac{S_d}{2\s}\int\limits_0^{+\infty}dk d\alpha_1 d\alpha_2 \frac{\alpha_1^{-z_2-1}k^{-2\ep/\s-z_2}}{\Gamma(-z_2)} \exp(-3\mu\alpha_1 -\alpha_1 k -\alpha_2\mu-\alpha_2 k)\eeq
After momenta integration we obtain
\beq I_k  = \frac{S_d\Gamma(1-2\ep/\s-z_2)}{2\s\Gamma(-z_2)}\int\limits_0^{+\infty}d\alpha_1d\alpha_2 (\alpha_1+\alpha_2)^{2\ep/\s+z_2-1}\alpha_1^{-z_2-1} \exp(-3\mu\alpha_1  -\alpha_2\mu)\eeq
First we will take care the integral over $\alpha_2$. We do substitution $\tilde\alpha_2=\alpha_1+\alpha_2$
\begin{align} I_k  &= \frac{S_d\Gamma(1-2\ep/\s-z_2)}{2\s\Gamma(-z_2)} \int\limits_0^{+\infty}d\alpha_1 \alpha_1^{-z_2-1} e^{-2\mu\alpha_1} \int\limits_{\alpha_1}^{+\infty} d\tilde\alpha_2 \tilde\alpha_2^{2\ep/\s+z_2-1}   e^{-\tilde\alpha_2\mu}
\nonumber\\
&= \frac{S_d\Gamma(1-2\ep/\s-z_2)\mu^{-z_2-2\ep/\s}}{2\s\Gamma(-z_2)}\int\limits_0^{+\infty}d\alpha_1 \alpha_1^{-z_2-1} e^{-2\mu\alpha_1}
\Gamma(2\ep/\s+z_2,\alpha_1\mu) 
\,,
\end{align}
where $\Gamma(\lm,x)$ is incomplete gamma function. The value of the last integral can be found in \cite{GR}. The final result or $I_k$ reads 
\beq\label{Ik} I_k  = \frac{S_d}{2\s}\frac{\Gamma(1-2\ep/\s-z_2)}{\Gamma(1-z_2)}\Gamma(2\ep/z_2)\mu^{-2\ep/\s}3^{-2\ep/\s}
\,_2F_1(1,2\ep/\s,1-z_2,2/3)\eeq
Inserting (\ref{Iq}) and (\ref{Ik}) into (\ref{I2zz}) we infer
\begin{align}  I_2 &= \frac{S_d\pi^{d/2}}{2\s(2\pi)^d} \frac{\Gamma(\s/2-\ep/2)}{\Gamma(-\ep/\s-1)^2} \mu^{-2\ep/\s}3^{-2\ep/\s}\Gamma(2\ep/\s)
\int\limits_{-i\infty}^{+i\infty}\frac{dz_1dz_2}{(2\pi i)^2} \,_2F_1(1,2\ep/\s,1-z_2,2/3) \frac{\Gamma(1-2\ep/\s-z_2)}{-z_2}
\nonumber\\
& \frac{\Gamma(1+z_1)}{\Gamma(\s(1+z_1)/2)} \frac{\Gamma(-z_1+z_2)}{\Gamma(\s(-z_1+z_2)/2)}
\frac{\Gamma(\ep/2+\s z_2/2)}{\Gamma(\s/2-\ep-\s z_2/2)} \Gamma(-\ep/2-\s z_1/2)\Gamma(-\s(-z_1+z_2)/2 -\ep/2+\s/2)\,.
\end{align}

First we sum over all poles of $\Gamma(-\ep/2-\s z_1/2)$ and then over pole at $z_2=0$. The result reads 
\begin{align}  I_2 &= \frac{S_d\pi^{d/2}}{2\s(2\pi)^d} \mu^{-2\ep/\s}3^{-2\ep/\s}\Gamma(2\ep/\s)
 \,_2F_1(1,2\ep/\s,1,2/3) \frac{\Gamma(1-2\ep/\s)}{\Gamma(\s/2-\ep)}\Gamma(\ep/2)
\nonumber\\
& \sum\limits^{+\infty}_{n=0}\frac{(-1)^n}{n!}\frac{\Gamma(1-\ep/\s+2n/\s)}{\Gamma(\s(1-\ep/\s+2n/\s)/2)} \frac{\Gamma(\ep/\s-2n/\s)}{\Gamma(\s(\ep/\s-2n/\s)/2)}
\Gamma(-\ep+n \s/2)\,.
\end{align}

 We will look the final result in the form 
\beq I_2 =   \mu^{-2\ep/\s}  e^{-B\ep} (c_{-2}\ep^{-2} +c_{-1}\ep^{-1}  ) \eeq
To obtain the divergent part of $I_2$ it is convenient to use {\tt MATHEMATICA}. The result for coefficients $c_{-2}$ and $c_{-1}$ are given by the formula (\ref{I2ep}).

\section{Appendix C}
Here we present the derivation of formula (\ref{SPexp}). Expanding scaling function $\ln(Z)$ at two-loop order and  \cite{Cardy} one  can infer that
\begin{align}  \ln(Z)&= \sum\limits_{1\le i < j \le p} \lm_{ij} n_in_j I_1 -\frac{1}{2}\left(\sum\limits_{1\le i < j \le p} \lm_{ij} n_in_j I_1\right)^2 -\sum\limits_{1\le i < j \le p} \lm^2_{ij} n_in_j I^2_1 -\frac{1}{2}\sum\limits_{1\le i < j \le p} \lm^2_{ij} n^2_in^2_j I^2_1 \nonumber\\
& - \sum\limits_{1\le i < j<k<l \le p} (\lm_{ij}\lm_{kl} + \lm_{ik}\lm_{jl}+\lm_{il}\lm_{jk}) n_in_jn_kn_l I^2_1 
+\frac{1}{2} \sum\limits_{1\le i < j \le p} \lm^2_{ij} n_in_j( n_i+n_j -2) I^2_1 \nonumber\\
& + \sum\limits_{1\le i < j<k \le p} (\lm_{ij}\lm_{ik} + \lm_{ij}\lm_{jk}+\lm_{ik}\lm_{jk}) n_in_jn_k I^2_1 -
2\sum\limits_{1\le i < j<k \le p} (\lm_{ij}\lm_{ik} + \lm_{ij}\lm_{jk}+\lm_{ik}\lm_{jk}) n_in_jn_k I_2 \nonumber\\
& - \sum\limits_{1\le i < j<k \le p} (\lm_{ij}\lm_{ik}n^2_in_jn_k + \lm_{ij}\lm_{jk}n_in^2_jn_k+\lm_{ik}\lm_{jk}n_in_jn^2_k)  I^2_1 \nonumber\\
& -\sum\limits_{1\le i < j \le p} \lm^2_{ij} n_in_j( n_i+n_j -2) I_2 +\frac{1}{2}\sum\limits_{1\le i < j \le p} \lm^2_{ij} n_in_j I^2_1
\,.
\end{align}
By the definition $\gm = \mu\frac{\pd\ln (Z)}{\pd\mu} $. After differentiation we use the formula 
$\lm_{ij}\mu^{-\ep/\s}=g_{Rij} +A g^2_{Rij}/\ep$, which one can infer from (\ref{lm-gR}), and the integral expansions (\ref{I2ep}) and \beq I_1^2 = \frac{2^{-2\s}\pi^{-\s}}{\ep^2\Gamma(\s/2)^2} + 
\frac{2^{-2\s}\pi^{-\s}}{\ep\Gamma(\s/2)^2}[\ln (4\pi) + \psi^{(0)}(\s/2)] + O(\ep^0)\eeq to derive the following result

\begin{align} \gm &= - \frac{1}{\s}\sum\limits_{1\le i < j \le p}  n_in_j g_{Rij} -\frac{1}{\ep\s}\sum\limits_{1\le i < j \le p}  n_in_j g^2_{Rij} + \frac{2}{\ep\s}\sum\limits_{1\le i < j \le p}  n_in_j g^2_{Rij} -\frac{1}{\ep\s}\sum\limits_{1\le i < j \le p}  n_in_j g^2_{Rij} 
\nonumber\\
& + \sum\limits_{1\le i < j<k \le p} n_in_jn_k (g_{Rij}g_{Rik} + g_{Rij}g_{Rjk}+g_{Rik}g_{Rjk}) 
\left(  \frac{2}{\s^2}\frac{2^{-2\s}\pi^{-\s}}{\Gamma(\s/2)^2}\ln(3/4)  \right)
\nonumber\\
& + \sum\limits_{1\le i < j \le p} g^2_{Rij} n_in_j( n_i+n_j -2) \left( \frac{1}{\s^2}\frac{2^{-2\s}\pi^{-\s}}{\Gamma(\s/2)^2}\ln(3/4)  \right)
\,.
\end{align}
The critical exponent is the value of this expression evaluated at the fixed point $g_{Rij}=\ep$. It easy to see that the resut is equivalent to (\ref{SPexp}).

\end{document}